# Dark Matter in an n-Space Expanding Universe

**Mario Rabinowitz**

Armor Research, 715 Lakemead Way, Redwood City, CA 94062-3922
E-mail:  Mario715@gmail.com

**Abstract**
The total number of degrees of freedom of a d-dimensional body in n-space is derived so that equipartition of energy may be applied to a possibly n-dimensional early universe.  A comparison is made of a range of proposals to include free and bound black holes as either a small component or the dominant constituent of dark matter in the universe.  The hypothesis that dark matter consists in part of atomic gravitationally bound primordial black holes is closely examined in 3-space, as well as in n-space; and the Chavda and Chavda holeum hypothesis is found to be flawed.  Blackbody and Hawking radiation are generalized to n-space, and Hawking radiation is shown to be simply proportional to the black hole density. The importance of quantum gravity in understanding the role of early universe dark matter is undermined because present approaches to a theory of quantum gravity violate the equivalence principle.  A general heuristic proof for geodesics on an expanding hypersphere is presented. Classical limits of Einstein's General Relativity are considered. A novel approach to the accelerated expansion of the universe is discussed.  An anomalous surprising aspect of 4-space is demonstrated.

# 1 Introduction

Black holes in differing scenarios have long been considered as candidates for dark matter. This paper will concentrate on gravitationally bound atoms (GBA) [6,14, 15], and free little black holes (LBH) as constituents of dark matter because LBH beamed radiation from LBH may account for the accelerated expansion of the universe. Because GBA have been proposed [6] to solve an important long-standing problem in astrophysics, GBA will be analyzed here in close detail.

Gravitational atoms have quantized orbits for the same reason that the orbits of ordinary electrostatic atoms are quantized. This follows directly from the quantization of angular momentum. The analysis is extended to macroscopic higher dimensions since some theories attribute physical reality to them, and a surprising result is found in 4-space. Dark matter may have played a critical role in the early universe that may have had more than 3 spatial dimensions. So let us deal with n-dimensional space in our considerations.

# 2 Equipartition of kinetic energy in n-space

### 2.1 Degrees of freedom in n-space

The total number of degrees of freedom $D_n$ of a d-dimensional body in n-space is

$$D_n = n + (n-1) + (n-2) + ... + (n-d), \qquad (2.1)$$

for d ≤ n. Once n coordinates establish the center of mass, there are (n - 1) coordinates left to determine a second reference point on the body, leaving (n - 2) for the third point, ..., and finally (n - d) coordinates for the (d + 1)th reference point.

Since the RHS of equation (2.1) has (d + 1) terms:

$$D_n = n + (n-1) + (n-2) + ... + (n-d) = (d+1)\left(\frac{n+(n-d)}{2}\right)$$
$$= \left(\frac{d+1}{2}\right)(2n-d). \qquad (2.2)$$

It is interesting to note that $D_n$ is the same for d = (n - 1) and for d = n:

$$D_n(d = n, or\ n-1) = \left(\frac{n+1}{2}\right)(2n-n) = \left(\frac{(n-1)+1}{2}\right)[2n-(n-1)] = \frac{n(n+1)}{2} \quad (2.3)$$

*2.2 Equipartition of kinetic energy*

In 3-space, $D_3$ varies from 3 for d = 0 (point-like object) to 6 for d = 2 (planar object like an ellipse) or d = 3 (object like a spheroid). Choosing n = 10 in reference to string theory, equation (2.2) shows that $D_{10}$ varies from 10 for d = 0 to 55 for d = 9 or 10.

Because the kinetic energy is a quadratic function of velocity in n-space, there will be on the average (1/2)kT of kinetic energy per degree of freedom $D_n$. Let us consider two cases: 1) 3-dimensional body (which could be bound by short range forces) in n-space, i.e. d = 3; 2) n-dimensional body in n-space, i.e. d = n, where n is the number of spatial dimensions in the space-time manifold of (n+1) dimensions.

For a 3-dimensional body in n-space, from equation (2.2) the average kinetic energy is

$$\langle KE \rangle = D_n \left(\tfrac{1}{2}kT\right) = \left(\frac{3+1}{2}\right)(2n-3)\left(\tfrac{1}{2}kT\right) = (2n-3)kT \qquad (2.4)$$

= 17 kT for n = 10. [5 kT for a point-like body, depending on scale.]

For an n-dimensional body in n-space, using equation (2.4) gives

$$\langle KE \rangle = D_n \left(\tfrac{1}{2}kT\right) = \left(\frac{n+1}{2}\right)(2n-n)\left(\tfrac{1}{2}kT\right) = \left(\frac{n(n+1)}{4}\right)kT. \qquad (2.5)$$

= 3 kT for n = 3. [(3/2)kT for a point-like body, depending on scale.]
= (55/2)kT   28 kT  for n = 10.

Thus in terms of equipartition of energy, at a given temperature T, there can be significantly higher kinetic energy than expected in higher dimensions. Chavda and Chavda [6] are interested in the early universe when the temperature $T \gg T_b = mc^2/k$ where m is each black hole mass which makes up holeum. In 3-space, the average kinetic energy is between 3/2 and 3 $mc^2$ depending on the scale of interaction as to whether the black holes should be considered point-like or 3-dimensional in collisions. From equations (2.4) and (2.5), in 10-space, holeum (if also bound by short range forces) would be dissociated, since the average kinetic energy is as high as 5 to 28 $mc^2$.

# 3 Radiation

*3.1 Blackbody radiation in n-space*

Let us generalize Boltzmann's derivation of the blackbody radiation law. In n-space, the radiation pressure $P_n = \tfrac{1}{n} u_n$, where $u_n$ is the energy density. The internal energy $U_n = u_n V_n$, where $V_n$ is the n-volume. The thermodynamic relation for internal energy is

$$\frac{\partial}{\partial N_n}(U_n)_T = T\left(\frac{\partial P_n}{\partial T}\right)_V - P_n \Rightarrow \frac{\partial}{\partial N_n}(u_n V_n) = T\frac{\partial}{\partial T}\left(\frac{u_n}{n}\right) - \frac{u_n}{n}. \quad (3.1)$$

Equation (3.1) leads to

$$\frac{du_n}{u_n} = (n+1)\frac{dT}{T} \Rightarrow u_n \propto T^{n+1}.$$
(3.2)

Thus the n-dimensional equivalent of the Stefan-Boltzmann blackbody radiation law from equation (3.2) is

$$P_{BBn} \propto cu_n \propto T^{n+1}, \quad (3.3)$$

It is interesting to note that the dimensionality of macroscopic space can be determined by measuring the exponent of the blackbody radiation law. If energetically stable atoms (e.g. bound by additional short-range forces) could exist in (n > 3)-space, equation (3.3) says that for a given T, the collective blackbody radiation of these atoms emits considerably higher power than in 3-space.

*3.2 Hawking radiation in n-space*

The Hawking radiation power, $P_{SH}$, follows from the Stefan-Boltzmann blackbody radiation power/area law $\sigma T^4$ for black holes. For Hawking [8, 9]:

$$P_{SH} \approx 4\pi R_H^2 [\sigma T^4] = 4\pi\left(\frac{2GM}{c^2}\right)^2 \sigma\left[\frac{\hbar c^3}{4\pi k GM}\right]^4 = \frac{\hbar^4 c^8}{16\pi^3 k^4 G^2}\{\sigma\}\left[\frac{1}{M^2}\right] \quad (3.4)$$

where $\sigma$ is the Stefan-Boltzmann constant. To avoid the realm of quantum gravity, Hawking requires the black hole mass $M > M_{Planck}$.

Since Hawking radiation [8, 9] was derived as blackbody radiation from a black hole, using equation (3.3), $R_{Hn}$ and $T_n$ from [15], the Hawking power radiated in n-space for n ≥ 3:

$$P_{SHn} \propto [R_{Hn}]^{n-1}[T_n]^{n+1} \propto [M^{1/(n-2)}]^{n-1}[M^{-1/(n-2)}]^{n+1} \propto \frac{1}{M^{2/(n-2)}} \quad (3.5a)$$

$$\propto M^{-2} \text{ for } 3-space; \text{ and } \propto M^{-1/4} \text{ for } 10-space.$$

Though ordinary blackbody radiation is dramatically large $\propto T^{11}$ in 10-space, the mass dependency of Hawking radiation decreases for dimensions higher than 3 for LBH.

Rabinowitz [16] has shown that Hawking radiation power is also given by

$$P_{SH} = \rho g \hbar / 90 \tag{3.5b}$$

where $\rho$ is the density of the black hole.

*3.3 Gravitational tunneling radiation (GTR)*

Gravitational tunneling radiation (GTR) may be emitted from black holes in a process differing from that of Hawking radiation, $P_{SH}$, which has been undetected for three decades. Belinski [4], a noted authority in the field of general relativity, unequivocally concludes "the effect [Hawking radiation] does not exist." GTR is offered as an alternative to $P_{SH}$. In the gravitational tunneling model [14], beamed exhaust radiation tunnels out from a LBH with radiated power, $P_R$, due to the field of a second body, which lowers the LBH gravitational potential energy barrier and gives the barrier a finite width. Particles can escape by tunneling (as in field emission). This is similar to electric field emission of electrons from a metal by the application of an external field.

Although $P_R$ is of a different physical origin than Hawking radiation, we shall see that it is analytically of the same form, since $P_R \propto \Gamma P_{SH}$, where $\Gamma$ is the transmission probability approximately equal to the WKBJ tunneling probability $e^{-2\Delta g}$ for LBH. The tunneling power [12] radiated from a LBH for $r \gg R_H$ is:

$$P_R \approx \left[ \frac{\hbar c^6 \langle e^{-2\Delta g} \rangle}{16 \pi G^2} \right] \frac{1}{M^2} \sim \frac{\langle e^{-2\Delta g} \rangle}{M^2} \left[ 3.42 \times 10^{35} W \right], \tag{3.6}$$

where M in kg is the mass of the LBH. No correction for gravitational red shift needs to be made since the particles tunnel through the barrier without change in energy. The tunneling probability $e^{-2\Delta g}$ is usually $\ll 1$ and depends on parameters such as the width of the barrier, M, and the mass of the second body [14].

Hawking invoked blackbody radiation in the derivation of equation (3.4). But it was not invoked in the GTR derivation of equation (3.6).[14] Although $P_R$ and $P_{SH}$ represent different physical processes and appear quite disparate, the differences in the equations almost disappear if we substitute into equation (6.4) the value obtained for the Stefan-Boltzmann constant $\sigma$ by integrating the Planck distribution over all frequencies:

$$\sigma = \left\{ \frac{\pi^2 k^4}{60 \hbar^3 c^2} \right\}, \tag{3.7}$$

$$P_{SH} = \frac{\hbar^4 c^8}{16 p^3 k^4 G^2} \left\{ \frac{p^2 k^4}{60 \hbar^3 c^2} \right\} \left[ \frac{1}{M^2} \right] = \frac{\hbar c^6}{16 p G^2} \left\{ \frac{1}{60} \right\} \left[ \frac{1}{M^2} \right]. \tag{3.8}$$

Thus $\quad P_R = 60 \langle e^{-2\Delta g} \rangle P_{SH}$. $\tag{3.9}$

Rabinowitz [16] has shown that GTR power is also given by
$$P_R = 2 r g \hbar e^{-2\Delta g} / 3. \tag{3.10}$$

# 4 Quantized Gravitational Orbits in n-Space

Let us consider quantized non-relativistic gravitational circular orbits in n-space, where n = 3, 4, 5, ... ∞ is the number of spatial dimensions in the space-time manifold of (n+1) dimensions. They are an analog of electrostatic atomic orbitals. Ordinary matter does not have a density high enough to make such orbits achievable, but LBH do. For example, in 3-space, a $10^{-5}$ kg LBH with a radius of ~ $10^{-32}$ m has a density of ~ $10^{90}$ kg/m$^3$, whereas nucleon densities are only ~ $10^{18}$ kg/m$^3$.

The results of previous derivations [12, 15] for quantized circular gravitationally bound orbits for the orbital radius, orbital velocity, binding energy, and the discrete gravitational spectrum (identical in form to the electromagnetic spectrum of the hydrogen atom) are generalized here. These were previously derived for orbiting mass m << M. Since Chavda and Chavda [6] primarily deal with m = M, the prior results are presented here in terms of the reduced mass $\mu = mM/(m+M)$ to directly assess limitations on their holeum analysis. The results will be presented in n-space and 3-space. Since they are more general and more precise, they supplant the previous results.

The orbital radius in n-space is
$$r_n = \left[ \frac{j \hbar p^{\frac{n-2}{4}}}{[2 G_n M m m \Gamma(n/2)]^{1/2}} \right]^{\frac{2}{4-n}}, \tag{4.1}$$

where j = 1, 2, 3, ... is the principal quantum number. M and m are gravitationally bound masses. The n-space universal gravitational constant $G_n$ changes, in a way that is model dependent, from its 3-space value. The Gamma function $\Gamma(n) \equiv \int_0^\infty t^{n-1} e^{-t} dt$ for all n (integer and non-integer). When n is an integer, G(n) = (n-1)! $\hbar$ is (Planck's constant)/2$p$. For comparison with Chavda and Chavda [6], the masses may be considered black holes with M = m.

In 3-space, equation (2.1) yields

$$r_3 = \frac{j^2\hbar^2}{GMm\,m} = \frac{2j^2\hbar^2}{Gm^3} \quad \text{for M = m.} \tag{4.2}$$

The orbital velocity in n-space is

$$v_n = \left\{\left[\frac{2pG_nMm\Gamma\left(\frac{n}{2}\right)}{p^{n/2}}\right]\left[\frac{m^{\frac{n-3}{4-n}}\left[2G_nMm\Gamma\left(\frac{n}{2}\right)\right]^{\frac{1}{4-n}}}{(j\hbar)^{2/(4-n)}p^{(n-2)/2(4-n)}}\right]\right\}^{1/2}. \tag{4.3}$$

In 3-dimensions equation (4.3) gives

$$v_3 = \frac{GMm}{j\hbar} = \frac{Gm^2}{j\hbar} \quad \text{for M = m.} \tag{4.4}$$

Although $v_3$ is independent of m, $v_n$ is not independent of m for higher dimensions. In some holeum cases, $m \sim m_{Planck}$ was used [6], for which equation (2.4) gives $v_3$ ~ c (the speed of light), necessitating special relativity corrections in their analysis. Smaller masses are also used which reduce $v_3 < c$, but as shown in Sections 2 and 6, this leads to an insufficient binding energy for their chosen realm.

The n-space acceleration is

$$a_n = \frac{-2pG_nMm\Gamma\left(\frac{n}{2}\right)}{mp^{n/2}}\left[\frac{[2G_nMm\,m\Gamma(n/2)]^{1/2}}{j\hbar p^{(n-2)/4}}\right]^{\frac{2n-2}{4-n}}. \tag{4.5}$$

In 3-space, equation (4.5) yields

$$a_3 = \frac{-G^3M^3m^3\,m}{(j\hbar)^4} = \frac{-G^3M^3m^4}{2(j\hbar)^4} \quad \text{for M = m.} \tag{4.6}$$

It is interesting to note from equations (4.1) to (4.6) that the acceleration, as well as the orbital radius and velocity, are functions of the mass m in all dimensions as a result of quantization, even when $M \gg m$. Though the presence of m may seem to be an artifact of the Bohr-Sommerfeld condition, the same mass dependency and basically the same results are obtained from the Schroedinger equation. The failure of m to vanish indicates that quantum mechanics is inconsistent with the weak equivalence principle (WEP). Since the strong equivalence implies the WEP, a violation of the WEP implies a violation of the SEP. Rabinowitz showed both an indirect and a direct violation of the SEP by non-relativistic and relativistic Quantum Mechanics [17,18]. Since the SEP is the cornerstone of Einstein's General Relativity (EGR), a theory of quantum gravity may not be possible combining traditional EGR and present day QM. Classically for M >> m, these variables are independent of the orbiting mass, since m cancels out in accord with the equivalence principle.

In n-space, the total energy of a gravitationally bound atom is

$$E_n = \frac{(n-4)m^{(n-2)/(4-n)}}{n-2}\left[\frac{G_n Mm\Gamma\left(\frac{n}{2}\right)}{p^{(n-2)/2}}\right]\left[\frac{\left[2G_n Mm\Gamma\left(\frac{n}{2}\right)\right]^{(n-2)/(4-n)}}{(j\hbar)^{2(n-2)/(4-n)} p^{(n-2)^2/2(4-n)}}\right]. \quad (4.7)$$

In 3-space, equation (4.7) reduces to

$$E_3 = -\frac{G^2 M^2 m^2 m}{2 j^2 \hbar^2} = -\frac{G^2 m^5}{4 j^2 \hbar^2} \text{ for M = m.} \quad (4.8)$$

Note from equation (4.7) that all energy levels are ≥ 0 in 4 and higher dimensional space, yielding the result for circular orbits that orbiting bodies in gravitational atoms cannot be bound by energy constraints in higher dimensions, no matter how strong the gravitational attraction, unless short range forces also come into play. (Similarly, for electrostatically bound atoms since they have the same dependence on n.) This is relevant to string theory if short range forces or other constraints cannot be invoked to achieve stability when their extra dimensions are unfurled.

Mathematically this results from the leading factor [(n-4)/(n-2)] in the complicated quantized equation (4.7). Why n > 3 leads to $E_n$ ≥ 0, can be understood in simpler terms for circular orbits. For a long-range attractive force like gravity with M >> m

$$F_n = \frac{-2p G_n Mm\Gamma(n/2)}{p^{n/2} r_n^{n-1}} = \frac{-mv_n^2}{r_n} \Rightarrow \frac{1}{2} mv_n^2 = \frac{p G_n Mm\Gamma(n/2)}{p^{n/2} r_n^{n-2}}, \quad (4.9)$$

where the (n-1) exponent of r in $F_n$ results from Gauss' law in n-space, e.g. $F_3 = -GMm/r^2$ because the area of a sphere $\propto r^2$, since we live in a 3-dimensional macroscopic space. Substituting equation (2.9) into the equation for total energy

$$E_n = \frac{1}{2} mv_n^2 + \frac{-2p G_n Mm\Gamma(n/2)}{(n-2) p^{n/2} r_n^{n-2}} = \frac{p G_n Mm\Gamma(n/2)}{p^{n/2} r_n^{n-2}} + \frac{-2p G_n Mm\Gamma(n/2)}{(n-2) p^{n/2} r_n^{n-2}}$$

$$= \left[\frac{n-4}{n-2}\right] \frac{p G_n Mm\Gamma(n/2)}{p^{n/2} r_n^{n-2}} \geq 0 \text{ for } n > 3. \quad (4.10)$$

This result, with the same prefactor [(n-4)/(n-2)], applies both classically and quantum mechanically since quantization does not change the sign of the co-factor $\frac{p G_n Mm\Gamma(n/2)}{p^{n/2} r_n^{n-2}}$ >0, for positive masses or if both masses are negative.

The same results would be obtained for any other long-range force like the electrostatic force. Short-range forces like the nuclear force are not expected to give this result. The results here indicate Euclidean 4-space is singular in that $r_4$ is infinite, and though angular momentum, $L_G = m_4 r_4 = [2mG_4 Mm/p]^{1/2}$ remains finite, $L_G$ and $L_E$ [Electrostatic L] can't be quantized in the usual way because their dependence on $r_4$ vanishes. This and no binding energy for atoms for 4-space has consequences for the 4-space Kaluza-Klein unification of general relativity and electromagnetism, as well as for string theory. In other dimensions, dependence on $r_n$ allows the orbital radius to adjust in quantization of L. Quantization of L, without quantization of r, in 4-space for gravitational and electrostatic atoms leads to interesting results such as possible quantization of m etc.

In higher dimensional space the trajectories are generally neither cirlcular, nor elliptical, as the orbits become non-closed curves. Although only circular orbits have been considered, the more complicated central force problem where there is also a radial velocity, yields the same conclusion. Rather than considering $E_n > 0$, we must take into consideration the effective potential energy. The general case can be put in the form of a one-dimensional radial problem in terms of the effective potential energy of the system, $V_n' = V_n + L^2/2mr^2.$

(4.11)

$V_n(r)$ is the potential energy, and L is the angular momentum which remains constant because there are no torques in central force motion.

## 5 Viable Black Hole Atoms

Ordinary gravitational orbits are in the high quantum number, continuum classical limit. In considering GBA, black holes are ideal candidates for the observation of quantization effects [12, 15], since for small orbits very high density matter is necessary. Furthermore, "A little black hole can trap charge internally and/or externally. It could easily trap ~ 10 positive or negative charges externally and form a neutral or charged super-heavy atom-like structure"[14]. Moderately charged black holes could form electrostatically and gravitationally bound atoms. For the present let us consider only gravitational binding where the black hole mass M >> m, the orbiting mass. To avoid complications related to quantum gravity, m can be considered to be made of ordinary matter such as a nucleon or group of bound nucleons.

Newtonian gravity is generally valid for $r > 10\, R_H$ since the difference between Einstein's general relativity and Newtonian gravitation gets small in this region. (The black hole horizon or Schwarzschild, radius is $R_H = 2GM/c^2$, where M is the mass of a black hole and c is the speed of light.) This approximation should be classically valid for all scales since the $|potential\,energy|$.

$$|V| = \frac{G(M)\gamma m}{r} < \frac{G(R_H c^2/2G)\gamma m}{10 R_H} = \frac{\gamma m c^2}{20} \tag{5.1}$$

is scale independent, where $\gamma = (1 - v^2/c^2)^{-1/2}$. Thus it is necessary that $|V|$ be smaller than 1/20 of the rest energy of the orbiting body of mass m. We will operate in the realm of Newtonian gravity and thus require the orbital radius $r > 10\ R_H$. From equation (2.2) with $j = 1$ and $M \gg m$:

$$r = \frac{\hbar^2}{(GM)m^2} = \frac{\hbar^2}{(R_H c^2/2)m^2} \geq 10 R_H. \tag{5.2}$$

Solving equation (5.2)

$$R_H \leq \frac{\hbar}{\sqrt{5}\,mc} = \frac{\lambdabar_C}{\sqrt{5}}, \tag{5.3}$$

where $\lambdabar_C$ is the reduced Compton wavelength of the orbiting particle. So $r \geq 10\ R_H$ is equivalent to the quantum mechanical requirement $\lambdabar_C \geq \sqrt{5} R_H$.

Now let us find a relationship between M and m that satisfies $r \geq 10\ R_H$.

$$r = \frac{\hbar^2}{GMm^2} \geq 10 R_H = 10 \frac{2GM}{c^2}. \tag{5.4}$$

Equation (5.4) implies that

$$Mm \leq \frac{\hbar c}{\sqrt{20}\,G} = \frac{(M_{Planck})^2}{\sqrt{20}}. \tag{5.5}$$

For $M = m$, r is a factor of 2 larger and equation (5.4) would yield $M \leq M_{Planck}/\sqrt{10}$. This is why it would be impossible to also avoid the realm of quantum gravity if the two masses are equal as is primarily done in [6].

Now we determine the ground state orbital velocity v in general for any M and m that satisfy $r \geq 10\ R_H$ by substituting equation (5.5) into equation (4.4) for v.

$$v = \frac{G(Mm)}{\hbar} \leq \frac{G}{\hbar}\left(\frac{\hbar c}{\sqrt{20G}}\right) = \frac{c}{\sqrt{20}} = 0.224c. \tag{5.6}$$

So special relativity corrections need only be small, but in some cases $v \approx c$ in [6].

Substituting equation (5.6) for v into equation (2.8), the binding energy is

$$E = -\frac{m}{2}\left[\frac{GMm}{\hbar^2}\right]^2 = -\frac{m}{2}[v^2] = -\frac{m}{2}\left[\frac{c}{\sqrt{20}}\right]^2 = -\frac{mc^2}{40}. \tag{5.7}$$

A large range of M >> m can satisfy these equations. For a numerical example, let m = $m_{proton}$ = 1.67 x $10^{-27}$ kg. Equation (5.5) implies that M = 6.36 x $10^{10}$ kg, with $R_H$= 9.43 x $10^{-17}$m. Equation (5.7) gives a binding energy E = 3.76 x $10^{12}$ J = 23.5 MeV, with v = 6.72 x $10^7$ m/sec. We want the binding energy E >> kT, so T must be << 2.72 x $10^{11}$K. Although this is much less than the unification temperature $T_{unif}$ ~ $10^{29}$ K, and such atoms would not be stable in the very early universe, they could be formed at later times and would be stable over most of the age of the universe. This assumes negligible Hawking radiation [4, 12, 14].

## 6  Holeum instability

Chavda and Chavda [6] propose (p. 2928) that the black holes and the holeum are created, "When the temperature of the big bang universe is much greater than $T_b$ = $mc^2/k_B$, where m is the mass of a black hole and $k_B$ [k here] is the Boltzmann constant...."  Let us examine whether the binding energy is great enough to hold holeum together in this high temperature regime. The binding energy between the masses m and m is given by j = 1 in equation (4.8).

In order for the binding energy given by equation (4.8) to be large enough to hold the holeum atom together for high energy collisions in this regime, it is necessary that

$$E_{binding} = |E_{j=1}| = \frac{G^2 m^5}{4\hbar^2} \geq kT >> kT_b = mc^2, \qquad (6.1)$$

where kT >> $kT_b$ = $mc^2$ is given in [6], as quoted above. Equation (6.1) implies that

$$m >> \sqrt{2}\left(\frac{\hbar c}{G}\right)^{1/2} = \sqrt{2} m_{Planck}. \qquad (6.2)$$

Equation (6.2) says that masses >> the Planck mass are needed for holeum to be stable in this high temperature regime. This is incompatible with the position in [6, p. 2932] that they are dealing with black holes less than the Planck mass, "In this paper, we consider black holes in the mass range $10^3$ GeV/$c^2$ to $10^{15}$ GeV/$c^2$." This limits the black hole masses from $10^{-24}$ kg to $10^{-12}$ kg, compromising the stability of holeum by tens of orders of magnitude. Both for stability and to circumvent the need for a theory of quantum gravity, masses    2 x $10^{-8}$ kg = $m_{Planck}$ are required. But this brings in problems of too small an orbital radius as shown next.

A mass $fm_{Planck} = f(\hbar c/G)^{1/2}$, where f is a pure number can be substituted into equation (6.2) for j = 1 to ascertain the orbital radius, i.e. the separation of the two black holes for the ground state of holeum.

$$r_{j=1} = \frac{2\hbar^2}{G[fm_P]^3} = \frac{2\hbar^2}{G\left[f\left(\frac{2\hbar c}{G}\right)^{1/2}\right]^3} = \frac{2}{f^3 c}\left[\frac{\hbar G}{2c}\right]^{1/2}. \tag{6.3}$$

Let us compare this radius with the black hole $R_H = 2GM/c^2$ for $m = f(\hbar c/G)^{1/2}$,

$$\frac{r}{R_H} = \frac{2}{f^3 c}\left[\frac{\hbar G}{2c}\right]^{1/2}\left[\frac{c^2}{2Gf(2\hbar c/G)^{1/2}}\right] = \frac{1}{f^4}. \tag{6.4}$$

For a stable orbit, $f = \sqrt{2}$, as determined by equation (5.2). This implies that r = $R_H$/4 . This is inconsistent with their use of Newtonian gravity (NG) which requires r > 2 $R_H$ just to avoid collision between the orbiting black holes. In NG, for equal black hole masses, each LBH orbits at a radius r/2 about the center of mass of the atom. For r > 10$R_H$, NG requires $f < 1/10^{1/4} = 0.56$, but then the masses are each 0.56 $M_{Planck}$, requiring quantum gravity. For some cases they have 2$R_H$<2r < 10$R_H$, that is still not adequate.

Higher dimensional atoms will not alleviate this conundrum for the mass or the radius, as shown in Section 4. Sections 5 and 6 show that the way out of this problem is to have the mass M be a little black hole which is massive, yet with $R_H$ << r, and an ordinary orbiting mass m << $M_{LBH}$.

## 7 LBH: Dark matter and accelerated expansion of universe

*7.1 Accelerated expansion of the universe in 3-space*
It had long been taken for granted that the expansion of the universe is either at a constant rate, or decelerating due to the gravitational attraction of all the mass in it. So it came as quite a surprise in 1998 when two independent international groups of astrophysicists at Lawrence Berkeley National Lab [11] in the U.S., and Mount Strombo and Siding Spring Observatories [19] in Australia, using type Ia supernovae to gauge distances, discovered that the universe is accelerating in its expansion. One viable competing explanation is that accelerated expansion of the universe is due to radiation from little black holes (LBH) propelling them outward and gravitationally towing ordinary matter with them. Little black holes may be the dark matter/dark energy representing 95% of the mass of the universe. [13, 14, 15]. The conventional view is that 4 - 5 % is ordinary matter, 23 - 25 % is dark matter, and about 70 - 73 % is exotic unknown dark energy unrelated to matter.

Theorists were quick to coin the term "dark energy" in concert with the already existing conundrum of "dark matter." The two have been considered as separate entities. In my model, both are essentially the same, or due to the same source i.e. little black holes (LBH). My reason that LBH should be considered to be energy rather than matter is that they are too small to be easily detected, and hence look smooth like energy rather than lumpy like matter.

As 95% of the mass of the universe, LBH essentially hold the universe together gravitationally, and their directed radiation contributes to its accelerated expansion. That many LBH doing Hawking Radiation would fry the universe. Unlike LBH that Hawking radiate, my LBH are long-lived because in my model of Gravitational Tunneling Radiation (GTR) they radiate much, much less and the radiation is beamed

The discovery of the accelerated expansion of the universe was hailed as epoch making. So when the papers [11, 19] first came out, I scoured them to see what value of accelerated they had found. It was nowhere to be found. It is model dependent in EGR. Yet I would have expected them to give some numbers as related to different models as it would put this profound breakthrough into proper perspective. As a simple example in flat-space let us make a non-relativistic estimate of the average acceleration for the universe to reach its present radius starting as a point.

$$\langle a \rangle \sim \frac{2R}{T^2} \sim \frac{2[15 x 10^9 light - yr]}{[15 x 10^9 yr]^2} = \frac{2c}{15 x 10^9 yr} = \frac{2(3 x 10^8 m/\sec)}{5 x 10^{17} \sec} \sim 10^{-9} m/\sec^2. \quad (7.1)$$

It would be helpful to put the present acceleration into perspective relative to this miniscule amount. It just goes to show that an extremely small acceleration acting for an extremely long time can have a huge effect.

GTR is beamed between a black hole and a second body, and is attenuated by the tunneling probability $\langle e^{-2\Delta g} \rangle$ compared to $P_{SH}$. Two LBH may get quite close for maximum GTR. In this limit, there is a similarity between GTR and what is expected from the Hawking model. GTR produces a repulsive recoil force between two bodies due to the beamed emission between them. Thus if LBH are the dark matter, their gravitational tunneling radiation (GTR) may be the source of the dark energy that is causing an accelerated expansion of the universe.

In the orthodox view of EGR, it is not as if the Big Bang concept represents an explosion in a pre-existent space. Rather it is that the Universe grew (perhaps exponentially) from a tiny size because it is creating more space for itself. EGR argues that the distant stars are moving away from us (and each other) because new space-time is being created between us and them, not because of any force.

Observations of the early universe fluctuations seem to point to a flat universe -- at least in the part we can observe. As shown in Sec. 5, Newtonian gravity is generally valid for r > 10 $R_H$ since the difference between Einstein's general relativity and Newtonian gravitation gets small in this region. Since the universe appears to be Euclidean (flat), let us make a quasi-relativistic Newtonian calculation:

$$F = \frac{-\gamma GMm}{R^2 \gamma^2 \left[1 - \beta^2 \sin^2 \theta\right]^{3/2}} \Rightarrow F_{rad} = \frac{-GMm}{\gamma R^2} \text{ for } \theta = 0 \text{ or } \pi; \text{ or} \quad (7.2)$$

$$\Rightarrow F_\perp = \frac{-\gamma^2 GMm}{R^2} \text{ for } \theta = \pi/2, \quad (7.3)$$

where $\theta$ is the angle between the radial direction and the recession velocity v, M is the mass of the universe, m is the mass of a receding star or galaxy, $\beta = v/c$, and $\gamma = \left[1 - \beta^2\right]^{1/2}$. For illustration, R is the radius of the universe, and for very large $\gamma$ (at the edge of the universe) the radial gravitational attraction is greatly reduced inversely proportional to $\gamma$, and the gravitational attraction in the direction perpendicular to the radial is greatly increased proportional to $\gamma^2$. This simplified quasi-relativistic Newtonian result seems to at least qualitatively compatible with EGR. If two bodies are receding from each other at near the speed of light, and space-time curvature can only propagate at the speed of light, intuitively this implies a weaker attraction between the two bodies than when they are at a fixed distance apart.

Nigel Cook (private communication) takes the position that the gauge boson radiation [gravitons] carries less energy per unit time and contributes less effect when it becomes excessively red-shifted from very early times after the big bang, and that there is also a canceling effect related to the higher density of the early universe.

*7.2 Accelerated expansion of the universe in n-space*
We can generalize eq. (7.2) for the gravitational force in n-space as given by eq. (14) in [15], for an expanding universe:

$$F_n = \frac{-\gamma 2\pi G_n Mm \Gamma(n/2)}{\pi^{n/2} R^{n-1} \gamma^2 \left[1 - \beta^2 \sin^2 \theta\right]^{3/2}}, \quad (7.4)$$

where $G_n$ is the gravitational constant in n-space, and $\Gamma$ is the Gamma function.

The radial force is given by $\theta = 0 \text{ or } \pi$:

$$F_{n\,rad} = \frac{-2\boldsymbol{p}G_n Mm\Gamma(n/2)}{\boldsymbol{g}R^2}. \tag{7.5}$$

We see that the radial force is reduced by $\boldsymbol{g}$ in n-space as it is in 3-space.

The perpendicular force is given by $\boldsymbol{a} = \boldsymbol{p}/2$:

$$F_{n\perp} = \frac{-\boldsymbol{g}^2 2\boldsymbol{p}G_n Mm\Gamma(n/2)}{R^{n-1}}. \tag{7.6}$$

*7.3 Geodesics on an expanding hypersphere*

Rather than via force calculations such as given by eqs. (7.2) and (7.3), trajectories are determined in EGR by geodesics (shortest distance between two points in the space i.e. on a given surface) in 4-spacetime. Geodesics are found by writing the equation for the length of a curve, and then minimizing this length using Euler's equation (sometimes called Euler-Lagrange equations) derived from the variational calculus. This can get unwieldy on a hypersphere -- particularly an expanding hypersphere. So to gain an intuitive insight, let us start by first considering a general simple heuristic proof that the shortest path joining 2 points on the surface of a sphere is the shorter arc of the great circle joining the 2 points. Sec. 7.3 and Sec. 7.4 will be done in the spirit of the Wheeler and Feynman paper [22] in which the seminal concept of half-retarded and half-advanced potentials was introduced without any equations.

*7.3.1 The Shortest distance between 2 points on a Sphere*

1. Consider a great circle through the points a and b on a sphere, whose plane has a third point at the center of the sphere. This great circle is a circumference C of the sphere and hence there is no larger circle on the sphere. This great circle has small arc length D and large arc length E.

2. Consider another smaller circle of circumference c containing points a and b, whose center is not at the center of the sphere. This smaller circle has small arc length D' and large arc length E'.

3. Consider a straight line through the sphere from point a to point b.

4. For easy visualization of the relative lengths of D and D', the circle c may be rotated about the points a and b so that it lies in the same plane as circle C. (This is just an aid for visualization, and is not a necessary step.)

5. Arc D is closer to the straight line since circle C has the least curvature i.e. it has the largest radius of curvature. The larger radius arc between two points is the shortest path because as the radius gets larger the arc approaches a straight line.

6. Arc D is closer to being a straight line than arc D' which has a higher curvature. Hence arc D is shorter. Any circle that would be closer to the straight line than arc D, could not also go through points a and b.

7. Therefore the shorter arc of the great circle (arc D) is the shortest distance on the surface of the sphere going between points a and b.

8. A related problem is to find the Shortest Distance on a General Solid Made of Planar Surfaces, e.g. a tetrahedron. Open up all the faces between the two points so they lie on a plane. If you can draw a straight line between the two points in the plane and have it cross the faces without crossing a non-face region, it is the shortest distance.

*7.3.2 The Longest distance between 2 points on a Sphere*

1. In *7.2.1* above, we have shown that the shorter arc of the great circle (arc D) is the shortest distance between the two points a and b.
2. The circumference C is the largest possible non-oscillatory path on the sphere.
3. Therefore C - D = E is the longest path i.e. the longest non-oscillatory arc.

*7.3.3 The Shortest distance between 2 points on an expanding Hypersphere*

For a static hypersphere the argument is similar to *7.2.1* above. For an expanding hypersphere we note that whatever the rate of expansion, the straight line through the hypersphere remains the shortest path between points a and b. Therefore the expanding shorter arc of a great circle on the expanding hypersphere is the shortest path between a and b. Although these results are intuitive, n-dimensional space has its surprises. For example, the volume of a radius r, infinite dimensional sphere = 0 because the n-volume of an n-sphere relative to an n-cube of side = r, peaks $\approx$ 5-dimensional space. Thereafter for large n, the ratio of the n-sphere volume to the n-cube volume is a quickly diminishing fraction which $\rightarrow 0$ as $n \rightarrow \infty$.[15]

*7.4 Classical limits of Einstein's General Relativity*

Let us examine possible classical limits of Einstein's General Relativity (EGR) since it is our best tool for analyzing the universe. We first explore possible violation of the strong equivalence principle (SEP) upon which EGR is based. The SEP states that locally, gravitation is indistinguishable from an equivalently accelerating reference frame (and vice versa). The weak equivalence principle (WEP) states that inertial mass is equal to gravitational mass so that the trajectory of a freely falling mass m in an external gravitational field of a mass M, is independent of m.

Consider a gedanken experiment with three spherically symmetric masses in a straight line with mass M an equal distance between $m_2$ and $m_1$ with M >> $m_2$ > $m_1$, and equal radii for $m_2$ and $m_1$. (See [12] for a detailed analysis of many cases.) When let go, the three bodies accelerate toward their common center of mass. Since the center of mass of the system is between the centers of M and $m_2$, $m_2$ will have a shorter distance to fall toward the CM; and $m_1$ will have a longer distance to fall than $m_2$ to reach the CM. All three bodies must reach the CM at the same instant because the CM cannot move in the absence of an external force. (A similar argument can be made for two masses.) The same conclusion holds if the masses are not collinear. **Since the lightest mass $m_l$ has to go the farthest distance to reach the CM, it must go the fastest relative to the CM. This argument is general and holds for any attractive force.**

For a gravitational force this clearly violates the WEP, since this free fall motion is mass dependent. The SEP implies the WEP. So this violation of the WEP, implies a violation of the SEP because in logic if A implies B, (not B) implies (not A). **Since EGR is founded on the SEP, it is only approximately a many body theory, but applies rigorously only for the motion of light mass test bodies falling in the field of a source of one or more much heavier bodies.**

## 8 Different views of black holes as dark matter candidates

Discovery of the nature of dark matter will help to define what the universe is made of. It will reveal the invisible particles carrying the gravitational glue that holds the universe, galaxies, and clusters of galaxies together, and determines the curvature of space. We should not arbitrarily rule out the possibility that dark matter can occasionally manifest itself on earth. To give a broad perspective all views known to me of black holes as dark matter candidates are now presented.

*8.1 Large black holes: $10^{14}$ kg $\leq M_{BH} \leq 10^{36}$ kg*

A review article of 1984 [5 and references therein] presents the prevailing view of black holes as constituents of dark matter. The article considers only rather massive black holes as a possible component of dark matter: "A third cold DM [dark matter] candidate is black holes of mass $10^{-16}$ $M_{sun}$ $\leq M_{BH} \leq 10^6$ $M_{sun}$, the lower limit implied by the non-observation of g rays from black hole decay by Hawking radiation...." ($M_{sun} = 2 \times 10^{30}$ kg.)

*8.2 Medium black holes: $10^{12}$ kg $\leq M_{BH} \leq 10^{30}$ kg*

Trofimenko in 1990 [20] discussed the possibility that black holes up to the mass of the sun, $M_{sun}$, are involved in geophysical and astrophysical phenomena such as in stars, pulsars, and planets. Although he did not explicitly consider them as candidates for dark matter, for him they are "universal centres of all cosmic objects." That makes them such candidates implicitly. He was not concerned with the ramifications of LBH radiation, nor the time for LBH to devour their hosts. His lower mass limit of $10^{12}$ kg comes from the failure to detect Hawking radiation, and expected smallest primordial mass survival.

## 8.3 *Primordial black holes:* $M_{BH} \sim 10^{13}$ *kg*

Beginning in 1993, Alfonso-Faus [1] proposed "primordial black holes, massive particles about $10^{40}$ times the proton mass" [$10^{40}(10^{-27}$ kg$) = 10^{13}$ kg] as his dark matter candidate. He goes on to say that they do not radiate by Hawking radiation, but does not comment on how they radiate. Elsewhere [2] he asserts a radiation wavelength of $10^8$ cm from black holes that is the geometric mean between the radius of such a primordial black hole ($10^{-12}$ cm) and the radius of the universe ($10^{28}$ cm). With such a long wavelength, he concludes that they radiate, "about $10^{40}$ times lower " than in the Hawking model and hence "they would still be around....."

## 8.4. *Higher Dimensional Primordial Black Holes:* $10^{29}$ *kg* $\leq$ M $\leq$ $10^{34}$ *kg*

Argyres et al [3] examine primordial black holes (PBH) in higher compact dimensions. They conclude that for 6 extra compact dimensions (9-space), 0.1 solar mass PBH are dark matter candidates, but that this increases to $\sim 10^4$ solar masses if there are only 2 or 3 extra dimensions (5 to 6-space). Smaller PBH might be expected, since for them PBH radiation is almost entirely gravitons. In standard Hawking radiation from LBH, > MeV photons would dissociate big bang nucleosynthesis products, devastating the presently propitious predictions of light element abundances. They conclude, "The lightest black holes that can be present with any significant number density in our universe today are thus formed immediately after the epoch of inflationary reheating."

## 8.5 *Primordial little black holes:* $10^{-7}$ *kg* $\leq$ $M_{LBH}$ $\leq$ $10^{19}$ *kg*

Starting in 1998, Rabinowitz proposed that black holes radiate by GTR allowing primordial LBH to be regarded as candidates for the dark matter of the universe [13, 14, 15]. These were the smallest masses ($10^{-7}$ kg to $10^{19}$ kg) considered until 2002. Since GTR can be greatly attenuated compared with Hawking radiation, cf. Section 6.4, this has strong implications down to the smallest masses of LBH, whether the LBH are free or GBA. For Hawking [9], the smallest LBH that can survive to the present is M $\sim 10^{12}$ kg . Let us see what GTR predicts.

The evaporation rate for a black hole of mass M is $d(Mc^2)/dt = -P_R$, which gives the lifetime

$$t = \frac{16\boldsymbol{p}G^2}{3\hbar c^4 \langle e^{-2\Delta g} \rangle} [M^3].$$

(8.1)

This implies that the smallest mass that can survive up to a time t is

$$M_{small} = \left( \frac{3\hbar c^4 \langle e^{-2\Delta g} \rangle}{16\boldsymbol{p}G^2} \right)^{1/3} [t^{1/3}].$$  (8.2)

Primordial black holes with $M \gg M_{small}$ have not lost an appreciable fraction of their mass up to the present. Those with $M \ll M_{small}$ would have evaporated away long ago.

Thus the smallest mass that can survive within $\sim 10^{17}$ sec (age of our universe) is

$$M_{small} \geq 10^{12} \langle e^{-2\Delta g} \rangle kg.$$  (8.3)

Hawking's result [8, 9] of $10^{12}$ kg is obtained by setting $e^{-2\Delta g} = 1$ in eq. (8.3). Since $0 \leq e^{-2\Delta g} \leq 1$, an entire range of black hole masses much smaller than $10^{12}$ kg may have survived from the beginning of the universe to the present than permitted by Hawking's theory. Although it was inappropriate to propose long-lived LBH in days gone by [12, 14, 15], it seems fashionable [10] from 2006 on to call them long-lived compact objects rather than LBH.

For example, if the average tunneling probability $\langle e^{-2\Delta \gamma} \rangle \sim 10^{-45}$, then $M_{small} \sim 10^{-3}$ kg. For $M_{univ} \sim 10^{53}$ kg, $V_{univ} \sim 10^{79}$ m$^3$ (radius of 15 x$10^9$ light-year = 1.4 x $10^{26}$ m), the average density of such LBH would be 1 LBH per $10^{23}$m$^3$. The velocity of our local group of galaxies with respect to the microwave background (cosmic rest frame), $v_{LBH} \sim 6.2$ x $10^5$ m/sec [21], is a reasonable velocity for LBH with respect to the earth. This may make it possible to detect their incident flux $\sim (10^{-23}/m^3)(6.2$ x $10^5$ m/sec$) \sim 10^{-17}/m^2$sec on the earth. [15]

*8.6 Non-radiating holeum : $10^{-24}$ kg $\leq M_{BH} \leq 10^{-12}$ kg*

In 2002 Chavda and Chavda [6] introduced a novel proposal that gravitationally bound black holes will not Hawking radiate by analogy to the neutron. It appears from my analysis that stable holeum cannot exist in in 3-space, or in any higher

-19-

dimensions. Therefore whether or not such an object might Hawking radiate is a moot point.

The analogy between holeum and a bound neutron may not apply. A neutron in free space decays with a half-life of about 10.6 minutes. The neutron spontaneously decays into a proton, an electron, and an antineutrino. This is energetically possible because the neutron's rest mass is greater than that of the decay products. This difference in rest mass manifests itself in an energy release of $1.25 \times 10^{-13}$ J (0.782 MeV). The situation in a nucleus is complicated by many factors such as Fermi levels of the neutrons and the protons, etc. Neutrons do decay in nuclei that are beta emitters despite their relatively large binding energy which is typically 1 to $1.4 \times 10^{-12}$ J (6 to 8 MeV). Other than the interesting neutron analogy, they give no compelling reasons for the absence of Hawking radiation in black hole GBA. Sec. 5 details a number of reasons that their Holeum is unstable, one of which is their neglect of special relativity.

It is relevant to note that non-relativistic quantum mechanics and even the semi-classical Bohr-Sommerfeld equation give accurate energy levels for hydrogen despite being non-relativistic. This is because they neglect the serendipitously near-canceling effects of both relativity and spin. One is the relativistic increase of the electron's mass as its velocity increases near the proton. The other is the interaction of the electron's intrinsic magnetic moment with the Coulomb field of the proton. Since a neutral LBH has no magnetic moment, there are no canceling effects and one may expect a much less reliable result from a treatment that neglects special relativity.

## 9 Conclusion

Little black holes were shown to be viable candidates for the dark matter and dark energy of the universe. A novel model for the accelerated expansion of the universe was presented. An intuitive insight was provided for geodesics on an expanding hypersphere.

Orbits in n-space were analyzed to see if higher dimensions could enhance the stability of gravitationally bound atoms. Instead, it was found that orbiting bodies in higher dimensional gravitational atoms cannot be bound by energy constraints in higher dimensions (elliptical orbits would not change this conclusion). Similarly for electrostatically bound atoms. This is because there is no binding energy for $n > 3$, no matter how strong the coupling between the two bodies. So even in the early universe when the gravitational force is very high because all the fundamental forces have the same strength, in higher dimensions there would be no gravitational or electrostatic atoms unless short-range forces come into play. This has ramifications for both Kaluza-Klein theory and string theory. An anomalous surprising aspect of 4-space presents theoretical opportunities.

This paper questions the domains of validity in [6]. Aside from this issue, it was also shown that even with the high binding energy of holeum, it is not enough to remain stable in collisions for their stated condition that $kT \gg kT_b = mc^2$ in 3-space [6]. This was done by deriving the degrees of freedom of a d-dimensional body in n-space, and applying the equipartition of kinetic energy.

An incompatibilty between quantum mechanics and the weak equivalence principle was demonstrated in Section 2. Perhaps this can shed light on why attempts to develop a theory of quantum gravity have led to discrepancies and even contradictions.